\documentclass[%
 aip,
 amsmath,amssymb,
 reprint,%
]{revtex4-1}

\usepackage{graphicx}
\usepackage{dcolumn}
\usepackage{bm}

\usepackage[utf8]{inputenc}
\usepackage[T1]{fontenc}
\usepackage{mathptmx}
\usepackage{etoolbox}

\makeatletter
\def\@email#1#2{%
 \endgroup
 \patchcmd{\titleblock@produce}
  {\frontmatter@RRAPformat}
  {\frontmatter@RRAPformat{\produce@RRAP{*#1\href{mailto:#2}{#2}}}\frontmatter@RRAPformat}
  {}{}
}%
\makeatother
\begin{document}

\preprint{AIP/123-QED}

\title[Supercontinuum Generation by Saturated Second-Order Nonlinear Interactions]{Supercontinuum Generation by Saturated Second-Order Nonlinear Interactions}
\author{Marc Jankowski}
\email{marc.jankowski@ntt-research.com}
\affiliation{NTT Research Inc. Physics and Informatics Labs, 940 Stewart Drive, Sunnyvale, California}
\affiliation{Edward L. Ginzton Laboratory, Stanford University, Stanford, California}

\author{Carsten Langrock}
\affiliation{Edward L. Ginzton Laboratory, Stanford University, Stanford, California}

\author{Boris Desiatov}
\author{Marko Lon\v{c}ar}
\affiliation{John A. Paulson School of Engineering and Applied Sciences, Harvard University, Cambridge, Massachusetts}

\author{M. M. Fejer}%
\affiliation{Edward L. Ginzton Laboratory, Stanford University, Stanford, California}

\date{\today}

\begin{abstract}
We propose a new approach to supercontinuum generation and carrier-envelope-offset detection based on saturated second-order nonlinear interactions in dispersion-engineered nanowaveguides. The technique developed here broadens the interacting harmonics by forming stable bifurcations of the pulse envelopes due to an interplay between phase-mismatch and pump depletion. We first present an intuitive heuristic model for spectral broadening by second-harmonic generation of femtosecond pulses, and show that this model agrees well with experiments. Then, having established strong agreement between theory and experiment, we develop scaling laws that determine the energy required to generate an octave of bandwidth as a function of input pulse duration, device length, and input pulse chirp. These scaling laws suggest that future realization based on this approach could enable supercontinuum generation with orders of magnitude less energy than current state-of-the-art devices.
\end{abstract}

\maketitle

\section*{\label{sec:intro}Introduction}


The generation of low-noise supercontinua from mode-locked lasers is an increasingly important nonlinear process in modern optical systems, with applications spanning nearly every field of science~\cite{Br_s_2023}. Examples include spectroscopy~\cite{Picque2019,Ye2005,Guo_2020}, precision metrology~\cite{Diddams2010}, environmental monitoring~\cite{acp-19-4177-2019}, attoscience~\cite{Baltu_ka_2003,Krausz_2016}, signal processing~\cite{Xu_2021}, and optical coherence tomography~\cite{Ji_2021}. Almost all approaches to supercontinuum generation (SCG) rely on guided-wave devices with either pure or effective third-order ($\chi^{(3)}$) nonlinearities, which broaden the input bandwidth through self-phase modulation (SPM) and four-wave mixing. Typical devices require input pulse energies on the order of hundreds or thousands of picojoules~\cite{Dudley2006}, with state-of-the-art devices using the $P\propto L^{-1}$ scaling of the necessary power associated with $\chi^{(3)}$ processes (where $L$ is the length of the nonlinear waveguide) to realize a coherent octave of bandwidth with tens of picojoules~\cite{Mayer2015,Krger2020}. When used for carrier-envelope-offset (CEO) detection to produce a frequency comb, these devices are followed by a discrete second-harmonic generation (SHG) stage, which frequency doubles the long-wavelength portion of the spectrum to overlap with shorter wavelengths, thereby generating an f-2f beatnote.





Recent work has focused on realizing SCG in waveguides with second-order ($\chi^{(2)}$) nonlinearities. In these systems spectral broadening can be accompanied by harmonic generation and frequency mixing to simultaneously generate multiple overlapping combs, which allows for f-2f beatnotes to be detected from the output of a single chip~\cite{Phillips2011OL,Carlson2017,Okawachi2018,Hickstein2019,Yu2019,Okawachi2020, Rutledge2021}. Phase-mismatched $\chi^{(2)}$ processes such as second-harmonic generation (SHG) contribute to spectral broadening in these devices by operating in a cascaded limit, where a negligibly depleted pulse input at the fundamental generates a phase-mismatched second-harmonic~\cite{Phillips2011OE,Viotti2018,Zhou2012,Rutledge2021,Hickstein2017}. In the absence of pump depletion, back-action by the generated second harmonic on the fundamental can be modeled effectively as self-phase modulation, with resulting behaviors similar to $\chi^{(3)}$ systems~\cite{BacheThesis,Desalvo1992,Schiek1993,Conti2002,Stegeman1996}. In practice, the strength of these effective nonlinearities is limited by dispersive effects, such as the rate of temporal walk-off between interacting harmonics. By eliminating leading-order dispersive effects (thereby realizing a quasi-static interaction), phase-mismatched SHG in nanophotonic periodically-poled lithium niobate (PPLN) waveguides has been used to demonstrate SCG at the few-picojoule level~\cite{Jankowski2020}. While this result had previously been interpreted as having realized large effective $\chi^{(3)}$ nonlinearities, we will show in this paper that the broadening mechanisms in these dispersion-engineered waveguides are qualitatively different from any previous approach to SCG. This can be seen by noting that spectral broadening in these devices occurs in the saturated limit, which invalidates the assumption of an undepleted fundamental. The mechanisms responsible for spectral broadening by saturated nonlinear interaction are poorly understood, and we address these questions in this article.

\begin{figure*}[t!]
\centering
\includegraphics[width=\textwidth]{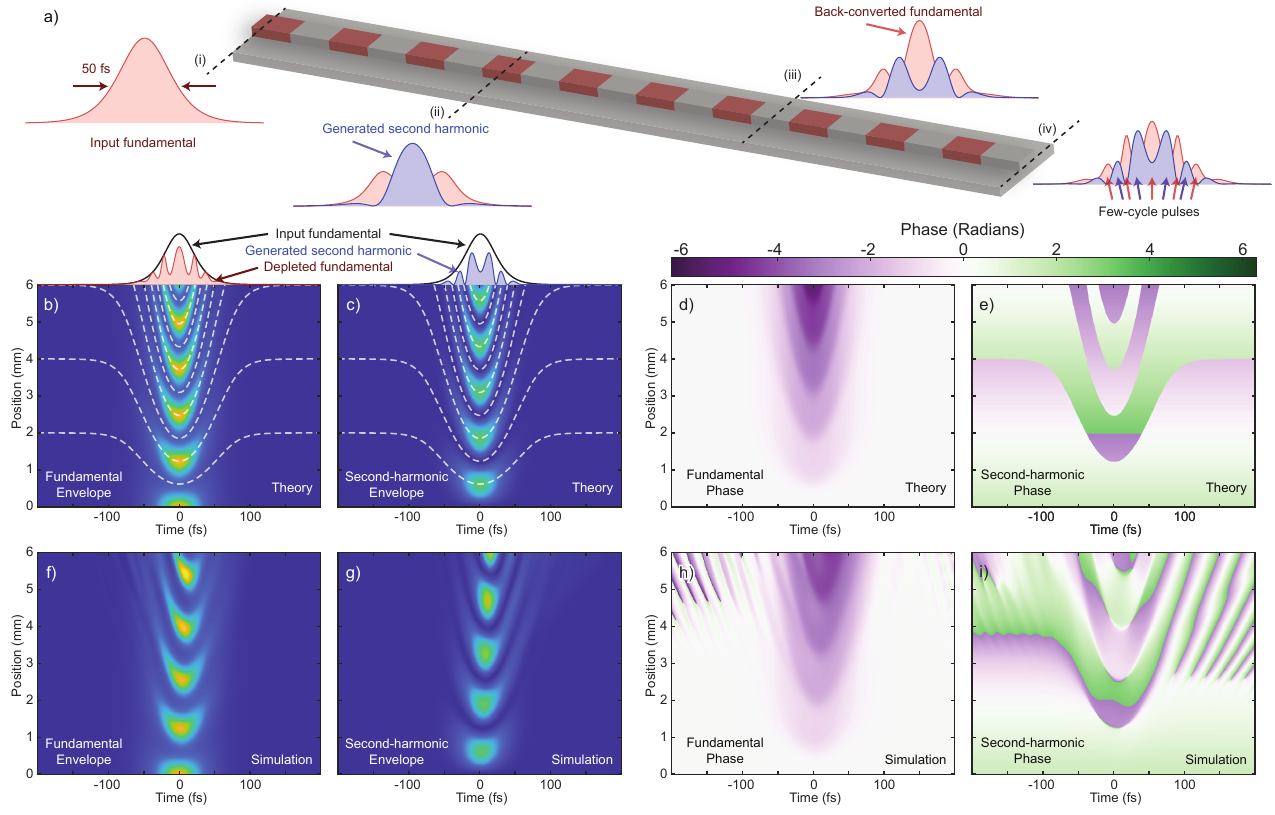}
\caption{\label{fig:jetheory} a) Schematic evolution of the two harmonics, $|A_\omega(t)|^2$ (red) and $|A_{2\omega}(t)|^2$ (blue) as each undergoes broadening. b,c) Theoretical evolution of $|A_\omega(t)|^2$ and $|A_{2\omega}(t)|^2$ based on Eqns.~\ref{eqn:je1}-\ref{eqn:je2}. Dashed white lines: conversion half-periods during propagation. d,e) The phase of the fundamental and second harmonic calculated using Eqns.~\ref{eqn:phase1}-\ref{eqn:phase2}. Both harmonics form plateaus of constant phase, which suggests spectral broadening is predominantly due to femtosecond amplitude substructure formed by pump depletion. f-i) Comparison of quasi-static theory with a split-step Fourier simulation including full dispersion relations and self-steepening for realistic waveguide parameters. The structure of each harmonic is largely unchanged by the small but non-zero second- and third-order dispersion. We note here that split-step Fourier simulations with no dispersion or self-steepening yield identical results to (b-e). The parameters used in (f-i) are chosen to compare the heuristic model of (b-e) to realistic device behaviors.}
\end{figure*}

In this work, we theoretically and experimentally investigate a new approach to SCG and CEO detection based on saturated $\chi^{(2)}$ interactions in dispersion-engineered QPM waveguides. This process relies on an interplay between saturated frequency conversion and phase-mismatch to quasi-periodically split the pulse envelopes associated with each harmonic into pulses with a shorter duration. When contrasted with traditional approaches to SCG based on either pure or effective $\chi^{(3)}$ nonlinearities this technique has many distinct features, including a different set of scaling laws and coherence properties. In principle, this approach may enable SCG with substantially lower energy requirements than processes based on $\chi^{(3)}$ nonlinearities due to both the relative strength of $\chi^{(2)}$ nonlinearities and the $P\propto L^{-2}$ power scaling associated with $\chi^{(2)}$ processes. Furthermore, when used for CEO detection, the f-2f beatnotes generated in the region of spectral overlap between each harmonic can remain in phase across hundreds of nanometers of bandwidth, as opposed to the tens of nanometers typically encountered in cascaded devices~\cite{Phillips2011OL}, which enables efficient CEO detection and eliminates the need for tunable narrow-band filters. The heuristic models studied here clarify the behaviors of SCG devices operating in the saturated limit, and the resulting scaling laws developed here provide a set of design rules for next generation SCG devices based on saturated $\chi^{(2)}$ interactions.

This paper proceeds in four sections. In sections \ref{sec:time} and \ref{sec:freq} we develop a heuristic model for SCG by saturated SHG in the time and frequency domains, respectively, and verify this model using split-step Fourier methods. While this model can be easily extended to three-wave mixing and optical parametric amplification we restrict our focus to SHG, which captures the essential physics. We then calculate the f-2f beatnotes in the region of spectral overlap between the two harmonics and show that these beatnotes can remain in phase across hundreds of nanometers of bandwidth. In section \ref{sec:expt} we compare this model to measurements of both the bandwidth and f-2f beatnotes generated by saturated SHG, which exhibit good agreement with the theoretical approach established in sections \ref{sec:time}-\ref{sec:freq}. While previous studies of supercontinuum generation in quasi-static waveguides interpreted spectral broadening as soliton compression due to an interplay between cascading and anomalous dispersion~\cite{Jankowski2020}, the measurements reported here occur in a normally dispersive regime where soliton compression cannot contribute to spectral broadening. Furthermore, the nonlinear length for SPM due to pure $\chi^{(3)}$ interactions can be shown to exceed the length of the waveguide for all input pulse energies studied here. Since neither of these processes can contribute to spectral broadening, this study represents the first unambiguous experimental demonstration of the regime proposed in sections \ref{sec:time}-\ref{sec:freq}. In section \ref{sec:scaling} we derive the scaling laws for the bandwidth generated by saturated $\chi^{(2)}$ interactions. This approach to SCG exhibits favorable length-scaling laws when compared to devices based on SPM, and may potentially realize SCG with hundreds of femtojoules in cm-scale devices. However, the energy requirements to achieve an octave of bandwidth scale with the cube of the input pulse duration, $\tau$, and therefore practical realizations require pulse durations shorter than $\sim$100 fs.



\section{Time-Domain Theory}\label{sec:time}

We consider the evolution of phase-mismatched fundamental and second-harmonic pulses in a dispersion-engineered quasi-phasematched waveguide in the limit where the field is sufficiently intense to deplete the fundamental and dispersion is negligible over the bandwidth of the pulses. This quasi-static model is motivated by several observations that suggest the cascaded nonlinearity model usually employed for $\chi^{(2)}$ SCG is inapplicable here: i) previous experimental demonstrations have shown that the generated supercontinua maintain coherence for soliton numbers far in excess of $\chi^{(3)}$ devices~\cite{Jankowski2020}, which suggests that the spectral broadening mechanisms may be different than an effective $\chi^{(3)}$; ii) the observed relative intensity of the second harmonic violates the assumption of an undepleted fundamental associated with the cascade regime; and iii), as will be shown in Sec.~\ref{sec:expt}, devices with saturated $\chi^{(2)}$ nonlinearities exhibit nearly identical behavior with either anomalous or normal dispersion, which suggests that soliton compression does not contribute meaningfully to spectral broadening.

The coupled-wave equations for the complex field envelopes $A_\omega(z,t)$ and $A_{2\omega}(z,t)$ are given by
\begin{subequations}
\begin{align}
\partial_z A_\omega &=-i\kappa A_{2\omega}A_\omega^* \exp(-i\Delta k z)-i\hat{D}_\omega A_{\omega},\label{eqn:CWE1}\\
\partial_z A_{2\omega} &= -i\kappa A_\omega^2 \exp(i\Delta k z) - \left(i\hat{D}_{2\omega}+\Delta k'\partial_t\right) A_{2\omega},\label{eqn:CWE2}
\end{align}
\end{subequations}
where $A_\omega$ is normalized such that $|A_\omega|^2$ is the instantaneous power of the fundamental wave. The temporal walk-off is given by $\Delta k' = v_{g,2\omega}^{-1}-v_{g,\omega}^{-1}$, where $v_{g,\omega}^{-1}=k'(\omega)$ is the inverse group velocity of the fundamental. The dispersion operators are given by $\hat{D}_\omega = \sum_{j=2}^{\infty}\left[(-i)^{j}k_\omega^{(j)}/j!\right]\partial^j_t$, where $k_\omega^{(j)}$ represents the $j^\mathrm{th}$ derivative of propagation constant $k$ at angular frequency $\omega$. Both $\Delta k'$ and $\hat{D}$ are assumed to be negligible in the quasi-static limit treated here. The appearance of $\Delta k'\partial_t$ in Eqn.~\ref{eqn:CWE2} is due to our choice of phase reference, which shifts the time coordinate to be co-moving with the group-velocity of the fundamental. $\Delta k = k_{2\omega}-2k_\omega-2\pi/\Lambda_G$ is the phase mismatch between the carrier frequencies of the interacting harmonics in the presence of a QPM grating with period $\Lambda_G$, and $\kappa=\sqrt{2\omega^2 d_\mathrm{eff}^2/\left(n^2_\omega n_{2\omega}\epsilon_0 c^3 \mathrm{A}_\mathrm{eff}\right)}\equiv\sqrt{\eta_0}$ is the nonlinear coupling, where $\eta_0$ is the conventional normalized SHG efficiency~\cite{Jankowski2020}. $A_\mathrm{eff}$ is the conventional effective area associated with $\chi^{(2)}$ nonlinear interactions~\cite{Jankowski2020}. In the quasi-static limit considered here ($\hat{D}_{\omega} = 0, \hat{D}_{2\omega} = 0$, $\Delta k'=0$), Eqns.\ref{eqn:CWE1}-\ref{eqn:CWE2} may be solved for the instantaneous field intensity at each point in time using the Jacobi elliptic functions associated with continuous-wave SHG in the limit of a depleted pump~\cite{Armstrong1962, Eckardt1984}. Defining the instantaneous field conversion efficiency as $|v(z,t)| = |A_{2\omega}(z,t)/A_\omega(0,t)|$, the field envelopes of the fundamental and second harmonic are 
\begin{subequations}
\begin{align}
A_\omega(z,t) = &\sqrt{1-v^2(z,t)}|A_\omega(0,t)|\exp(i\phi_\omega(z,t)),\label{eqn:je1}\\
A_{2\omega}(z,t) &= v(z,t)|A_\omega(0,t)|\exp(i\phi_{2\omega}(z,t))\label{eqn:je2},
\end{align}
\end{subequations}
where $v(z,t)=v_b(t) \mathrm{sn}\left(\kappa A_\omega(0,t)v_b^{-1}(t) z | v_b^4(t)\right)$. Here $\mathrm{sn}(u|m)$ is defined as the Jacobi elliptic sine, which continuously deforms from $\sin(u)$ to $\text{tanh}(u)$ as $m$ is increased from $0$ to $1$, and $v_b(t)=-|\Delta k/(4 \kappa A_\omega(0,t))|+\sqrt{1+|\Delta k/(4 \kappa A_\omega(0,t))|^2}$. We note here that $v_b^2(t)$ is the maximum pump depletion attainable for a given $\Delta k$ as a function of the local field amplitude input to the waveguide $A_\omega(0,t)$.


The Jacobi elliptic solutions found here bear many similarities to the sinusoidal evolution that occurs during phase-mismatched SHG in the weak-conversion limit, namely, periodic oscillations of the fundamental and second harmonic power in $z$, and a maximum pump depletion ($\nu_b^2$) that increases with $A_\omega(0,t)$. However, the Jacobi elliptic functions become increasingly anharmonic with increasing field intensity (due to saturation) and the spatial period at which power oscillates between the fundamental and second harmonic, hereafter referred to as the conversion period, decreases for a given $\Delta k$ as the local field intensity becomes larger. This conversion period is given by
\begin{equation}
L_\mathrm{conv}(t)=\frac{2 v_b(t) K(v_b^4(t))}{\kappa A_\omega(0,t)},
\end{equation}
where $K$ is the complete elliptic integral of the first kind. $K(v_b^4(t))$ varies slowly for most physically encountered values of $v_b(t)$, e.g. $K(v_b^4=0)=\pi/2$ and $K(v_b^4=0.81)\approx 1.45 \pi/2$ for a maximum pump depletion of $v_b^2 = 0.9$. Therefore the variation of $L_\mathrm{conv}(t)$ is dominated by $v_b(t)/(\kappa A_\omega(0,t))$. Fig. \ref{fig:jetheory}(a-c) shows the theoretical evolution of a 50-fs-wide (3 dB) sech$^2$ pulse in a 6-mm-long waveguide given by Eqns.~\ref{eqn:je1}-\ref{eqn:je2}. Here, we have assumed a pulse energy of 4 pJ, $\eta_0 = 1000~\%$/W-cm$^2$, and $\Delta k = -3\pi/L$, where $L$ is the length of the waveguide. The dotted white lines correspond to the $m^\mathrm{th}$ half-period, $L_m(t)= mL_\mathrm{conv}(t)/2$, where even $m$ coincide with the local maxima and minima of the fundamental and second harmonic, respectively. Near the peak of the pulse the conversion period is the shortest and both harmonics undergo $\sim 5$ conversion periods as the field propagates through the waveguide. The oscillations of the power in the tails of the pulse asymptotically approach to the conversion period associated with undepleted SHG (equal to twice the conventional coherence length in this limit), $L_\mathrm{conv}(\infty) = 2\pi/|\Delta k|$. Remarkably, the power at the peak oscillates three times faster than in the tails of the pulse, which generates a pulse shape with rapid femtosecond amplitude oscillations as each time-slice of the pulse cycles through a different number of conversion periods (Fig.~\ref{fig:jetheory}(a-c), red and blue curves).


Using the same quasi-static heuristic, the Jacobi elliptic solutions can be shown to predict phase envelopes for the fundamental and second harmonic,
\begin{subequations}
\begin{align}
\phi_\omega(z,t) - \phi_\omega(0,t) &= -\frac{\Delta k}{2}\int_0^z \frac{v^2(0,t)-v^2(z',t)}{1-v^2(z',t)}dz'\label{eqn:phase1}\\
&=\frac{1}{2}\sin^{-1}\left(\frac{\Delta k |A_{2\omega}(z,t)|}{2\kappa |A_{\omega}(z,t)|^2}\right)-\frac{\Delta k z}{4},\nonumber\\
\phi_{2\omega}(z,t) - 2\phi_\omega(0,t) &= - \pi/2 + \frac{\Delta k z}{2},\label{eqn:phase2}
\end{align}
\end{subequations}
respectively, and are plotted in Fig.~\ref{fig:jetheory}(d,e) for $\phi_\omega(0,t)=0$. We note here that the rate of phase accumulation has a fixed sign determined by $\Delta k$, and therefore in this context $\sin^{-1}(\sin(x))=x$ is defined to be a monotonic function. The rate of phase accumulation by the fundamental depends strongly on the degree of pump depletion, with large phase shifts accumulated at values of $z$ and $t$ that correspond to local maxima of $v(z,t)$. This behavior results in a saturable effective SPM for the fundamental, with the total accumulated phase plateauing across large time bins (Fig.~\ref{fig:jetheory}(d)). The phase envelope of the second harmonic is independent of time when the input fundamental is unchirped, $\phi_{2\omega}(z,t)=\phi_{2\omega}(z,0)$, and therefore can be neglected in the context of spectral broadening. The second-harmonic phase variation shown in Fig.~\ref{fig:jetheory}(d), $\angle A_{2\omega}(z,t)$, contains contributions from both $\phi_{2\omega}(z,t)$ and sign changes of $\nu(z,t)$, and therefore exhibits phase discontinuities of $\pm \pi$ every $L_\mathrm{conv}(t)$. These two behaviors, namely, the flat second-harmonic phase envelope and the plateaus of constant phase across the fundamental, suggest that the predominant broadening mechanism for saturated SHG is not effective SPM. Instead, the spectral broadening of each harmonic is dominated by the femtosecond-scale intensity variations imparted by pump depletion, as is verified by comparing the Fourier spectrum of the pulse including the full time-dependent phase of the fundamental to that when the fundamental phase is artificially set to zero.

We now verify this model against a split-step Fourier simulation assuming realistic parameters for dispersion-engineered waveguides, namely, a temporal walk-off of $\Delta k'=$5 fs/mm, group velocity dispersion for the fundamental and second harmonic of $k_\omega''=$9.5 fs$^2$/mm and $k_{2\omega}''=$70 fs$^2$/mm, respectively, and third-order dispersion given by $k_\omega'''=$-1100 fs$^3$/mm and $k_{2\omega}'''=$1200 fs$^3$/mm. These parameters were chosen to correspond to the dispersion relations of the TE$_{00}$ waveguide modes of the waveguides studied in Sec.~\ref{sec:expt}. The time-domain instantaneous power associated with each envelope, $|A_\omega|^2$ and $|A_{2\omega}|^2$, is shown in Fig. \ref{fig:jetheory}(f-g), respectively. The phase associated with each envelope is shown in Fig. \ref{fig:jetheory}(h-i). We note here that while the fundamental phase envelope is unwrapped to better visualize the phase accumulated during propagation, a similar procedure cannot be applied to the second harmonic due to the phase discontinuities accumulated around $L_\mathrm{conv}(t)$. To facilitate comparisons between theory and simulation we have left the second-harmonic phase wrapped. While the simulated pulse envelopes exhibit some distortion due to second- and third-order dispersion, the key aspects of our Jacobi elliptic approach such as the rapid amplitude variations of the resulting pulses are largely preserved. Given this strong agreement, we now develop a heuristic for the spectral broadening of each harmonic.

\begin{figure}[t]
\centering
\includegraphics[width=\columnwidth]{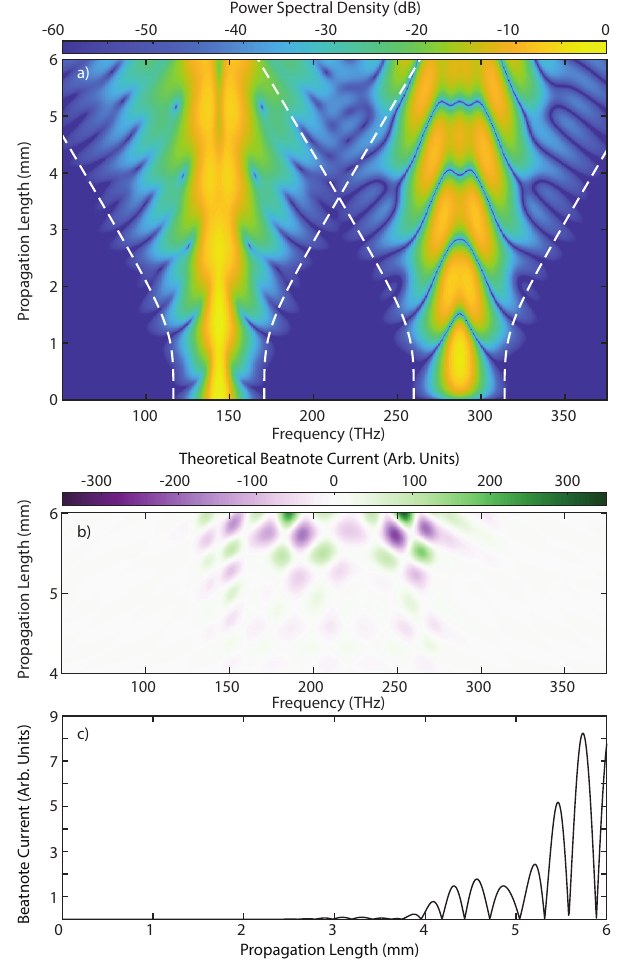}
\caption{\label{fig:jepsd}a) Theoretical power spectral density associated with the field envelopes shown in Fig. \ref{fig:jetheory}. b) Photocurrent spectral density for the f-2f beatnote as a function of optical frequency. The beatnotes fall in- and out-of-phase during propagation, with some positions producing beatnotes that remain in phase across 10's of THz of bandwidth. c) The resulting detected photocurrent, $I_\text{beat}(z)$, oscillates for increasing propagation length due to the quasi-periodic re-phasing shown in (b).
}
\end{figure}

\section{Frequency-Domain Theory}\label{sec:freq}

Having calculated both the amplitude and phase of each envelope, we can now Fourier transform these envelopes to study the evolution of the power spectral density. While we cannot obtain closed-form expressions for the Fourier-domain fields from Eqns. 2 and 4, we may gain several insights from the time-domain model that qualitatively capture the behavior of the generated spectrum. First, as discussed previously, the phase of the two harmonics has a negligible contribution to spectral broadening, with $\phi_\omega$ forming plateaus of nearly constant phase and $\phi_{2\omega}(z,t)=\phi_{2\omega}(z,0)$ contributing no time-dependent phase modulation. Second, we note that for $z>L_\mathrm{conv}(0)$ the number of local maxima contained in the instantaneous power of each envelope grows linearly with the number of half-periods around $t=0$. Finally, we note that each envelope loses two local maxima for every conversion period in the tails of the pulse ($L_\mathrm{conv}(\infty)$). Therefore, the number of local maxima for each harmonic is given by $N_\omega(z)\approx 2z\left[ L_\mathrm{conv}^{-1}(0)-L_\mathrm{conv}^{-1}(\infty)\right]$ and $N_{2\omega}(z)=N_\omega(z) + 1$ for $z > L_\mathrm{conv}(0)$. Both harmonics have one local maximum for $z < L_\mathrm{conv}(0)$. Since the instantaneous power $|A_{\omega}(z,t)|^2+|A_{2\omega}(z,t)|^2$ is conserved, each field envelope effectively splits into $N$ pulses with a duration $\sim \tau/N$, where $\tau$ is the pulse duration input to the waveguide (e.g. $\tau_\mathrm{FWHM}=1.76\tau$ for a sech pulse). Based on these observations we expect three behaviors in the frequency domain: i) a bandwidth $\Delta \nu$ that grows linearly for $z$ that satisfy $N_\omega(z)>2$ (or $z\gg L_\mathrm{conv}(0)$), ii) a constant bandwidth for $z$ that satisfy $N_\omega(z)<2$, and iii) the appearance of fringes in the frequency domain that become more finely patterned with increasing $z$, due to interference of $N$ the pulses formed in the time-domain.

The combined power spectral density associated with each harmonic, $|\hat{A}_\omega(z,\nu)|^2+|\hat{A}_{2\omega}(z,\nu)|^2$ is plotted in Fig.~\ref{fig:jepsd}(a), for the parameters used in Fig.~\ref{fig:jetheory}. The white dotted lines in Fig.~\ref{fig:jepsd}(a) correspond to a heuristic formula for the spectral width based on the arguments made above,
\begin{equation}
\Delta \nu_\omega(z) = \Delta \nu(0)\left(1 + (m N_\omega(z))^p\right)^{1/p},\label{eqn:je_bandwidth}
\end{equation}
where $p=4$ is a constant chosen to capture the transition from constant to linearly growing bandwidth around $z\approx L_\mathrm{conv}(0)$ and $m\approx 0.52 + 0.12\log(|\kappa A_\omega(0,0)/\Delta k|)$ is a numerically determined slope that corresponds to the rate change of bandwidth with respect to the number of generated pulses. The slope $m$ is a slowly varying function of $\kappa A_\omega(0,0)/\Delta k$ since the Jacobi elliptic functions become increasingly anharmonic with larger field intensity. This behavior reduces the temporal extent of each pulse formed between the zeroes of the Eqns. \ref{eqn:je1}-\ref{eqn:je2}, thereby altering the rate of the generated bandwidth with respect to the number of pulses formed. The characteristic input bandwidth $\Delta \nu(0)=7.6/(\pi^2\tau)$ is chosen to correspond to the $-60$ dB level. As can be seen from Fig.~\ref{fig:jepsd}(a), Eqn.~\ref{eqn:je_bandwidth} (dashed white lines) captures the spectral broadening due to quasi-static $\chi^{(2)}$ interactions for most cases of interest, and is valid for $2|\kappa A_\omega(0,0)|>|\Delta k|$. For $2|\kappa A_\omega(0,0)|<|\Delta k|$ the dynamics undergo a transition from saturated SHG to a cascaded nonlinearity, and the generated bandwidth correspondingly exhibits a sub-linear scaling with $z$. The power spectral density shown in Fig.~\ref{fig:jepsd}(a) exhibits interference fringes that become more finely patterned with increasing $z$, as expected from the qualitative picture discussed previously. The harmonics merge at the -60 dB level for $\Delta \nu_\omega(z) > \omega/(4\pi)$, or $z > 4$~mm in Fig.~\ref{fig:jepsd}(a), which enables f-2f interferometry in the region of spectral overlap.

The beatnote power contained in each spectral bin is calculated using $i_\mathrm{beat}(z,\nu)\propto 2\mathrm{Re}\left(\hat{A}_\omega(z,\nu)\hat{A}_{2\omega}^*(z,\nu)\right)$. Fig.~\ref{fig:jepsd}(b) shows the calculated $f_\mathrm{CEO}$ beatnote current as a function of optical frequency for the overlapping spectra. The beatnotes fall in and out of phase during propagation and, remarkably, for a suitable choice of power or device length the $f_\mathrm{CEO}$ beatnotes can remain in phase across nearly a micron of bandwidth (from 150 to 300 THz). As shown in Fig.~\ref{fig:jepsd}(c), this quasi-periodic rephasing of the beatnotes causes the total beatnote current obtained by integrating over the full bandwidth, $I_\text{beat}(z)=\int_0^\infty i_\text{beat}(z,\nu) d\nu$, to exhibit oscillations in $z$. In practice, the pulse energy used to drive the waveguide can be chosen to align a local maximum of the beatnote current with the length of the device. This process simplifies CEO detection by allowing the output of the waveguide to be focused on a photoreceiver with minimal filtering, while also improving the detected beatnote current by integrating the photocurrent over many comb lines.


\begin{figure}[t]
    \centering
    \includegraphics[width=\columnwidth]{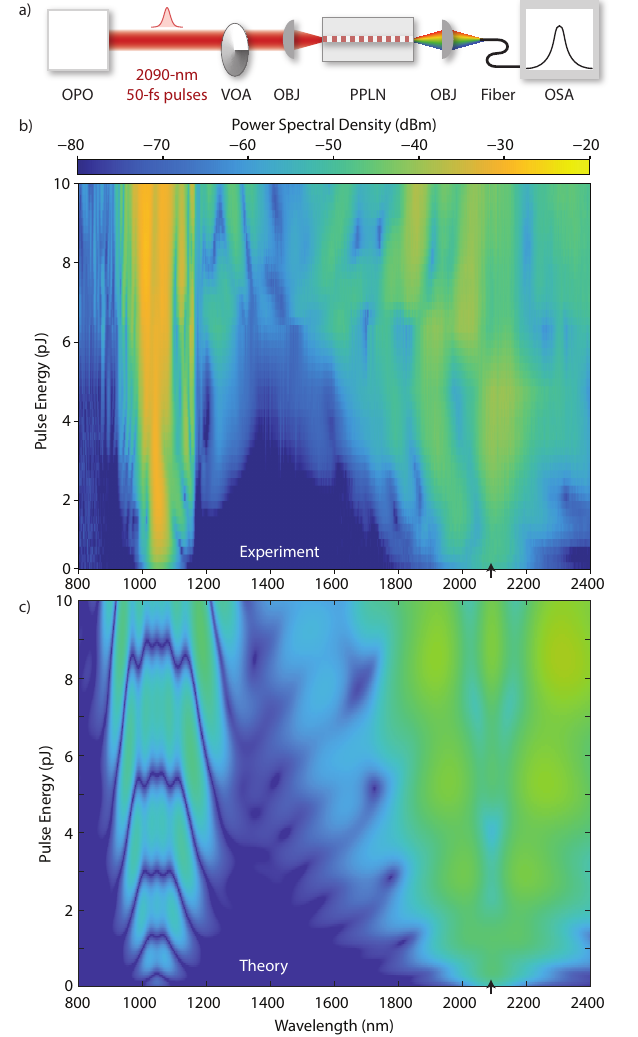}
    \caption{a) Experimental Setup. (OPO) Optical parametric oscillator, (VOA) variable optical attenuator, (OBJ) metallic Cassegrain objective, (OSA) optical spectrum analyzer. b) The experimentally measured and, c) theoretically calculated power spectral density as a function of input pulse energy. The black arrows centered around 2090 nanometers in (b) and (c) denote the pump wavelength.}
    \label{fig:Fig3}
\end{figure}

\section{Experimental Results}\label{sec:expt}

Having established the physical processes responsible for spectral broadening during saturated SHG, we now compare this model to supercontinua produced by the nanophotonic PPLN devices studied in~\cite{Jankowski2020}, including here new data obtained to more rigorously test model predictions. The design and fabrication of these devices were reported previously in~\cite{Zhang2017,Wang2018,Jankowski2020}, and we summarize the relevant aspects here. The waveguides under study have a film thickness of 700 nm, a top width of $\sim 1850$ nm, and an etch depth of $\sim$ 340 nm, which results in quasi-static operation around 2090 nm. Fifteen copies of these waveguides are poled with periods ranging from 5.01 $\mu$m to 5.15 $\mu$m; this step size of 10 nm corresponds to a shift of the accumulated phase mismatch $\Delta k L$ by $4.6 \pi$. Fine tuning of the phase-mismatch may be performed by changing the temperature of the waveguide, and phase-matched SHG is observed around a period of 5.11 $\mu$m and a temperature of 50 C. For this experiment, we use an adjacent waveguide with a poling period of 5.10 $\mu$m and operate at room temperature, which corresponds to a phase-mismatch of $\Delta k L \sim -3\pi$.

The experimental setup is shown in Fig. \ref{fig:Fig3}(a). As in previous studies~\cite{Jankowski2020}, the waveguides are driven using 50-fs-long pulses produced with a repetition rate of 100 MHz from a synchronously pumped degenerate optical parametric oscillator. We note, however, that in contrast with previous studies the pump pulses are centered around 2090 nanometers, instead of 2060 nanometers. This choice of wavelength renders the group velocity disperison 6 fs$^2$/mm, instead of -15 fs$^2$/mm, which eliminates any spectral broadening due to soliton self-compression. These pulses are coupled into the PPLN waveguides using a reflective inverse-cassegrain objective (Thorlabs LMM-40X-P01) to ensure that the in-coupled pulses and collected harmonics are free of chromatic aberrations. We record the output spectrum from the waveguide using two spectrometers: the near-infrared (600 - 1600 nm) is captured with a Yokogawa AQ6370C, and the mid-infrared (1600 - 2400 nm) is captured using a Yokogawa AQ6375. The results are shown in Fig.~\ref{fig:Fig3}(b). The fundamental and second harmonic are observed to broaden for input pulse energies in excess of 100 fJ, with the two harmonic merging at the -40 dB level for pulse energies as low as 4-pJ. This observed broadening with increasing pulse energy is consistent with the quasi-static theory shown in Fig.~\ref{fig:Fig3}(c). Furthermore, we observe a number of qualitative similarities between the spectra observed in theory and experiment. In particular, for pulse energies between one to five picojoules, the power spectrum of the fundamental exhibits a local minimum around the carrier frequency of the fundamental, 2090 nanometers. For pulse energies greater than five picojoules, this local minimum splits into two minima centered symmetrically around the carrier frequency, with a local maximum at 2090 nanometers. Similar patterns occur in the tails of the spectra; the spectrum of the fundamental forms successive local minima and maxima in the band between 1600 - 1800 nm with increasing pulse energy, and the second harmonic exhibits oscillatory tails between 1200 - 1400 nm. The experimentally observed second harmonic is, however, much brighter than the envelope predicted by the heuristics developed here. The origin of this effect is under investigation.

\begin{figure}[t!]
    \centering
    \includegraphics[width=\columnwidth]{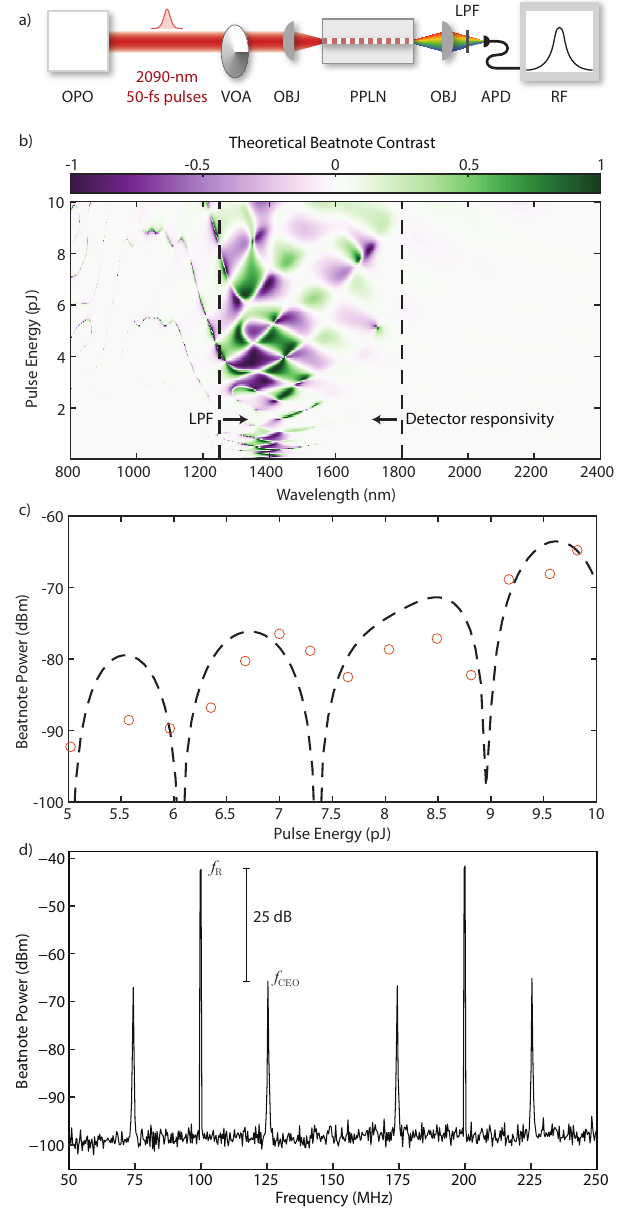}
    \caption{a) Experimental Setup. (LPF) Long-pass filter, (APD) avalanche photodiode, (RF) radio-frequency spectrum analyzer. b) Theoretical beatnote contrast as a function of wavelength and pulse energy. We detect the range from 1250 - 1800 nanometers (dashed black lines) using a long-pass filter and an InGaAs avalanche photodiode. c) Measured beatnote power as a function of pulse energy. Orange circles correspond to experiment, and the dashed black line corresponds to theory. d) Measured f-2f beatnotes, for a pulse energy of $\sim 10$ picojoules.}
    \label{fig:Fig4}
\end{figure}

For energies $>$5 pJ, f-2f beatnotes may be detected in the region of spectral overlap by filtering the light output from the waveguide and focusing it onto a photodiode (Fig. \ref{fig:Fig4}(a)). Figure \ref{fig:Fig4}(b) shows the beatnote contrast predicted by the quasi-static theory, $2\mathrm{Re}\left(\hat{A}_\omega(L,\lambda)\hat{A}_{2\omega}^*(L,\lambda)\right)/\left(|\hat{A}_\omega(L,\lambda)|^2+|\hat{A}_{2\omega}(L,\lambda)|^2\right)$, as a function of wavelength and pulse energy. The contrast is highest in the spectral region from 1200 - 1800 nanometers. Therefore, we detect f-2f beatnotes by filtering the light output from the waveguide using a Thorlabs FELH1250 longpass filter, and focus this light on a MenloSystems APD310 avalanche photodiode. We record the RF beatnotes using a Rigol DSA815 RF spectrum analyzer with the resolution bandwidth set to 10 kHz. The RF beatnote power in dBm is shown in Fig. \ref{fig:Fig4}(c) as a function of input pulse energy. We observe oscillations of the beatnote power with increasing pulse energy, which qualitatively agrees with the periodic re-phasing of the beatnotes predicted by theory (Fig. \ref{fig:Fig4}(c), dashed line). The beatnote current achieves a local maximum corresponding to an RF power of -65 dBm around an input pulse energy of 9.75 picojoules. Fig. \ref{fig:Fig4}(d) compares the relative intensity of the measured $f_\mathrm{CEO}$ beatnote with the $f_R$ beatnote corresponding to the repetition rate of the pulses for a pulse energy of 9.75 pJ. The detected $f_\mathrm{CEO}$ beatnote power is only 25 dB below the $f_R$ beatnote, even when the detected optical bandwidth spans $\sim$550 nm.

\section{Scalability of this approach}\label{sec:scaling}

In experimental realizations of SCG, reductions of the required pump pulse energy are typically realized either by using shorter pulses or by fabricating longer devices. In this section, we derive these scaling laws for the power required to produce a coherent octave by saturated $\chi^{(2)}$ nonlinearities, and compare them to conventional SCG devices. The three scaling laws considered here are summarized in Table~\ref{tab:scaling laws}. We first consider length rescalings given by $z\rightarrow s_1 z$. The solutions to Eqns.~\ref{eqn:CWE1}-\ref{eqn:CWE2} are scale invariant with a simultaneous rescaling of the device length, dispersion orders, and field intensity given by $z\rightarrow s_1 z$, $\Delta k\rightarrow\Delta k/s_1$, $\Delta k'\rightarrow\Delta k'/s_1$, $\hat{D}_\omega\rightarrow\hat{D}_\omega/s_1$, $\hat{D}_{2\omega}\rightarrow\hat{D}_{2\omega}/s_1$, and $P_\mathrm{in}(t)=|A_\omega(0,t)|^2+|A_{2\omega}(0,t)|^2 \rightarrow P_\mathrm{in}(t)/s_1^2$. Therefore, in the quasi-static limit an increase of device length by a factor $s_1$ correspondingly results in a quadratic reduction of the pulse energy necessary to achieve the same degree of spectral broadening at the output, $U_\mathrm{in}\rightarrow U_\mathrm{in}/s_1^2$, provided that the dispersion of the waveguide remains sufficiently negligible over the length of the device. In contrast, the nonlinear Schr\"{o}dinger equation used to describe SCG in waveguides with $\chi^{(3)}$ nonlinearities is scale invariant when $z\rightarrow s z$, $\hat{D}\rightarrow\hat{D}/s$, and $P_\mathrm{in}(t)=\rightarrow P_\mathrm{in}(t)/s$, which exhibits a linear reduction of the energy required to produce a supercontinuum as the length of the waveguide is increased. While state-of-the-art devices based on $\chi^{(3)}$ nonlinearities have been able to achieve SCG with tens of picojoules using long waveguides~\cite{Mayer2015,Krger2020}, the quadratic scaling of the energy requirements associated with a $\chi^{(2)}$ process would enable octave-spanning SCG with hundreds of femtojoules of pulse energy by rescaling the waveguide designs shown here to centimeters.

\begin{figure}[t!]
    \centering
    \includegraphics[width=\columnwidth]{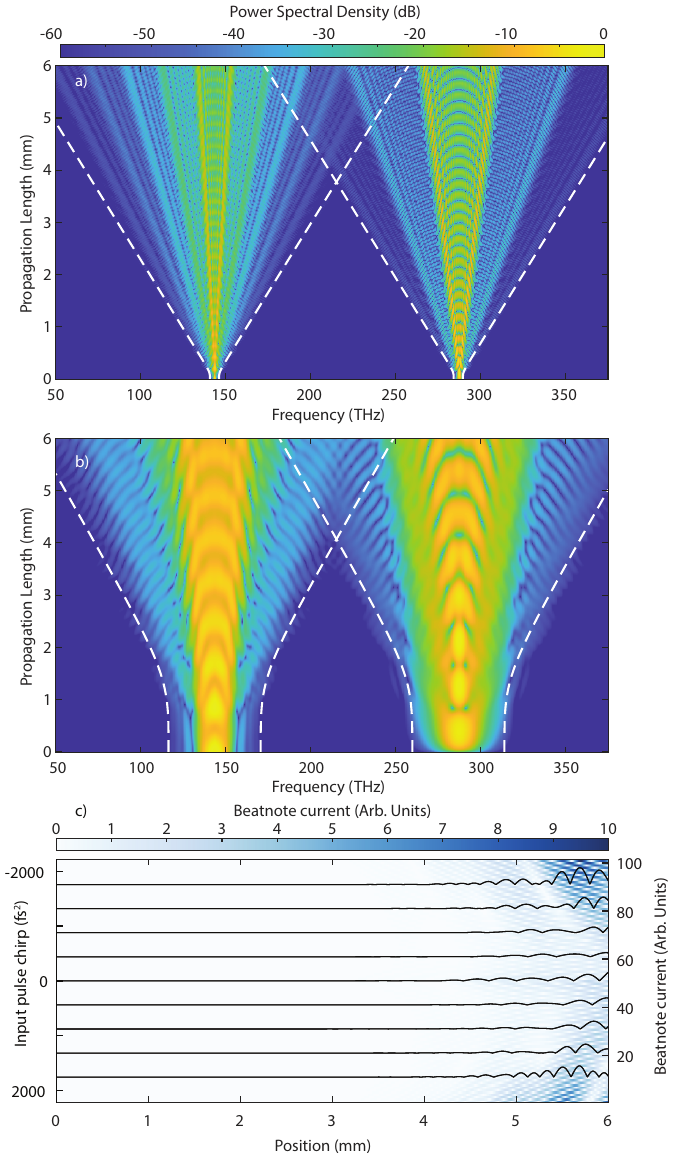}
    \caption{Bandwidth-invariant rescalings of the dynamics shown in Fig.~\ref{fig:jepsd}. Spectral evolution for a) a transform-limited pulse with $s_2 = 10$ ($U_\text{in}=4$~nJ and $1.76\tau=500$~fs), and b) a chirped pulse with $s_2=2$ ($1.76\tau=100$~fs, $1.76\tau_0=50$~fs, and $U_\text{in}=32$~pJ). Dashed white lines correspond to Eqn.~\ref{eqn:je_bandwidth2}. c) Beatnote current as a function of input pulse chirp, $\phi''$, with the input power rescaled by $(\tau/\tau_0)^3$. Values of $\phi''$ are chosen to realize $\tau_0 \leq \tau \leq 2\tau_0$. Solid black lines: $I_\text{beat}(z)$ (in arb. units) for coarsely sampled $\phi''$, showing similar integrated beatnote currents for all values of $\phi''$.}
    \label{fig:Fig5}
\end{figure}

We now consider the role of pulse duration by using Eqn. \ref{eqn:je_bandwidth}. In the limit of large nonlinear coupling ($\kappa A_\omega(0,0)\gg \Delta k$), or large $z$, Eqn.~\ref{eqn:je_bandwidth} can be approximated as $\Delta \nu(z) = \Delta \nu(0)\kappa A_\omega(0,0)z/\pi$. This expression suggests an \emph{approximate} scaling law for the generated bandwidth as a function of the input pulse duration $\tau$. Increasing $\tau$ by a factor $s_2$ reduces both the input bandwidth and intensity, $\Delta \nu(0)\rightarrow \Delta \nu(0)/s_2$ and $A_\omega(0,0)\rightarrow A_\omega(0,0)/\sqrt{s_2}$, which increases the power required to achieve a desired $\Delta \nu$ as $U_\mathrm{in}\rightarrow U_\mathrm{in}s_2^3$. An example of this bandwidth-invariant rescaling is shown in Fig.~\ref{fig:Fig5}(a). Here, a 4-nJ, 500-fs ($s_2 = 10$) pulse achieves a similar amount of bandwidth as the original spectral evolution shown in Fig.~\ref{fig:jepsd}(a). This cubic scaling restricts us to pulse durations on the order of $\lesssim$~100 fs simply because the energy requirements of longer pulses are impractical. Even a 200-fs-long pulse would require hundreds of picojoules of pulse energy to achieve the output bandwidths demonstrated here, comparable to the pulse energies that have already been used successfully to generate octave-spanning $\chi^{(3)}$ SCG in LN nanowaveguides~\cite{Yu2019,Lu2019}. While these limitations impose rather strict requirements on the input pulse duration, this rapid scaling suggests that these devices may work well when combined with soliton compression~\cite{Foster_2005,Carlson_2019}.

Given this strong power scaling with input pulse duration, a natural concern for experimental realizations is the degradation of generated bandwidth due to chirped input pulses. This case is readily analyzed using Eqn.~\ref{eqn:je1}-\ref{eqn:phase2} provided that the input field and phase envelopes, $A_\omega(0,t)$ and $\phi_\omega(0,t)$, are calculated using $A_\omega(0,t) = \int_{-\infty}^{\infty}d\nu A_\omega(0,\nu')\exp(-i\phi'' (2\pi\nu')^2/2)\exp(-2\pi i \nu' t)$, where $\nu'$ is the frequency offset from $\omega$ and $\phi''$ is the group delay dispersion of the input pulse. Closed form solutions for $A_\omega(0,t)$ are well known for Gaussian pulses. In the case of a sech pulse, the time-domain envelope can be approximated using an ansatz $A_\omega(0,t) = \sqrt{U/(2\tau)}\mathrm{sech}(t/\tau)\exp(-ib t^2)$, where the pulse duration $\tau = \tau_0(1 + (2\phi''/(\pi\tau_0^2))^2)^{1/2}$ and the chirp parameter $2b=\phi''((\phi'')^2+(\pi\tau_0^2/2)^2)^{-1}$ are calculated using Lagrangian methods~\cite{Malomed2002}. The transform-limited pulse duration attained when $\phi''=0$ is $\tau_0$.

\begin{table}

\begin{tabular}{p{3.2 cm}|c|c|c|c|c|p{3 cm}}
Scaling law & $L$ & $\Delta k$ & $\tau_0$ & $\tau$ & $U_\text{in}$ & Notes\\
\hline
\hline
Length rescaling & $s_1$ & $s_1^{-1}$ & 1 & 1 & $s_1^{-2}$ & Fully scale invariant\\
Pulse duration rescaling & 1 & 1 & $s_2$ & $s_2$ & $s_2^{3}$ & $\Delta f$ invariant\\
Chirp & 1 & 1 & 1 & $s_3$ & $s_3^{3}$ & $\Delta f$ invariant
\end{tabular}
\caption{\label{tab:scaling laws} Scaling laws for spectral broadening by saturated $\chi^{(2)}$ interactions in quasi-static waveguides. Quadratic reductions of the required input energy can be achieved by scaling to longer interaction lengths. Conversely, using longer a pulse duration at the input results in cubic increase of the required pulse energy.}
\end{table}

Since the phase envelopes $\phi_\omega(z,t)$ and $\phi_{2\omega}(z,t)$ do not contribute meaningfully to spectral broadening for $z\gg L_\mathrm{conv}(0)$, we expect the bandwidth for large $z$ to be well described by our quasi-static model for a pulse with duration $\tau$. Conversely, for small $z$, the bandwidth will be given by the transform-limit of a pulse with duration $\tau_0$. Based on these observations, we may generalize Eqn.~\ref{eqn:je_bandwidth} to account for chirp
\begin{equation}
\Delta \nu_\omega(z) = \Delta \nu(0)\left(1 + \left(mN_\omega(z)\tau_0/\tau\right)^p\right)^{1/p},\label{eqn:je_bandwidth2}
\end{equation}
where $N_\omega(z)$ is calculated using the peak instantaneous power of the chirped pulse and $\Delta \nu(0)$ is calculated using the transform-limit pulse duration $\tau_0$. This heuristic is shown to agree well with power spectra calculated using Eqns.~\ref{eqn:je1}-\ref{eqn:phase2} in Fig.~\ref{fig:Fig5}(b). Equation~\ref{eqn:je_bandwidth2} suggests that the scaling laws for chirp are identical to those of pulse duration rescalings ($\tau \rightarrow s_3 \tau_0$, $U_\text{in}\rightarrow s_3^3 U_\text{in}$), provided that the chirped pulse duration $\tau$, rather than the transform-limited duration $\tau_0$, is used to calculate the increase in energy requirements. $I_\text{beat}(z)$ is shown as function of $\phi''$ in Fig.~\ref{fig:Fig5}(c). The solid black lines are coarse samples of $I_\text{beat}(z)$, offset to intercept the y-axis at the corresponding value of $\phi''$. All traces are seen to exhibit hard zeroes and comparable heights in the overlap region ($z>4$~mm), which suggests that the integrated current is robust with respect to chirp when the input pulse energy is simultaneously rescaled by $s_3^3$. These behaviors suggest that only the pulse duration $\tau$, rather than the chirp parameter $b$, must be well controlled to achieve broadband f-2f beatnotes with realistic pulse energies. Notably, $b$ is typically measured using sophisticated techniques such as FROG, and grows linearly for small values of input dispersion ($b\approx \phi''/(\pi\tau_0^2)$ when $2|\phi|'' < \pi\tau_0^2$). In contrast, $\tau$ is easily measured using standard autocorrelation techniques and is a weak function of $\phi''$ when $2|\phi''| < \pi\tau_0^2$. Since only $\tau$ needs to be well controlled, we expect any deleterious effects due to pulse chirp to be easily managed in further experimental realizations of $\chi^{(2)}$ SCG.


\section{Conclusion}

We have established a theoretical model of supercontinuum generation based on saturated quasi-static $\chi^{(2)}$ interactions, and have demonstrated the first experimental verification of this model by studying spectral broadening in PPLN nanowaveguides. In contrast with the effective self-phase modulation that occurs with cascaded $\chi^{(2)}$ interactions, here spectral broadening occurs due to a femtosecond amplitude substructure that forms across the pulse envelope of the fundamental and second harmonic. This process generates coherent octaves of bandwidth with picojoules of pulse energy and produces f-2f beatnotes that can remain in phase across hundreds of nanometers of bandwidth. These broadband beatnotes simplify f-2f detection and improve the signal-to-noise ratio of the detected $f_\mathrm{CEO}$ beatnote since the detected photocurrent can be integrated over many comb lines. Finally, we use our model to derive a set of scaling laws that provide simple design rules for devices based on saturated $\chi^{(2)}$ interactions. These scaling laws suggest that $\chi^{(2)}$ supercontinua may access substantially lower energy scales than state-of-the-art $\chi^{(3)}$ devices both by using longer nonlinear sections and shorter input pulses.

\begin{acknowledgments}
The authors wish to thank NTT Research for their financial and technical support. Electrode patterning and poling was performed at the Stanford Nanofabrication Facility, the Stanford Nano Shared Facilities (NSF award ECCS-2026822), and the Cell Sciences Imaging Facility (NCRR award S10RR02557401). Patterning and dry etching was performed at the Harvard University Center for Nanoscale Systems (CNS), a member of the National Nanotechnology Coordinated Infrastructure (NNCI) supported by the National Science Foundation.
\end{acknowledgments}

\section*{Funding}
National Science Foundation (NSF) (ECCS-1609549, ECCS-1609688, EFMA-1741651, CCF-1918549, OMA-2137723); AFOSR MURI (FA9550-14-1-0389); Army Research Laboratory (ARL) (W911NF-15-2-0060, W911NF-18-1-0285).

\section*{Data Availability Statement}

The data that support the findings of this study are available from the corresponding author upon reasonable request.

\section*{Author Declaration Statement}

The authors have no conflicts to disclose.

\bibliography{biblio}

\end{document}